# Effect of Mn Substitution on Superconductivity in PrFeAs(O,F): Role of Magnetic Impurities

Priya Singh[1], Konrad Kwatek[2], Tatiana Zajarniuk[3], Taras Palasyuk[2], Cezariusz Jastrzębski[2], A. Szewczyk[3], Michał Wierzbicki[2], Shiv J. Singh[1*]

[1]*Institute of High Pressure Physics (IHPP), Polish Academy of Sciences, Sokołowska 29/37, 01-142 Warsaw, Poland*

[2]*Faculty of Physics, Warsaw University of Technology, Koszykowa 75, 00-662, Warsaw, Poland*

[3]*Institute of Physics, Polish Academy of Sciences, Aleja Lotnikow 32/46, 02-668 Warsaw, Poland*

**\*Corresponding author:**

Email: sjs@unipress.waw.pl

https://orcid.org/0000-0001-5769-1787

# Abstract

We investigate Mn substitution at the Fe site in $PrFe_{1-x}Mn_xAsO_{0.7}F_{0.3}$ ($0 \leq x \leq 0.1$) using structural, Raman, density functional theory (DFT), transport, and magnetic measurements. X-ray diffraction and Raman analyses confirm preferential Mn incorporation into the FeAs planes, accompanied by lattice expansion and suppression of Fe-related vibrational modes. Electrical transport reveals a systematic decrease of the superconducting transition temperature from 48 K ($x = 0$) to complete suppression at $x = 0.1$, together with low-temperature resistivity upturns evolving toward insulating-like behavior. Magnetization and magnetotransport measurements show degradation of superconducting coherence, critical current density, upper critical field, and vortex activation energy with increasing Mn content. The results demonstrate that Mn acts as an efficient magnetic impurity, strongly perturbing the electronic and magnetic environment of the FeAs layers. Comparative analysis indicates relatively enhanced robustness of superconductivity in the Pr-based system, highlighting the role of rare-earth-dependent electronic correlations in impurity effects.

# I. INTRODUCTION

The discovery of superconductivity at the critical transition temperature $T_c\sim26$ K in LaFeAsO$_{1-x}$F$_x$ [1], initiated intensive research into iron-based superconductors (IBS) as a platform for studying unconventional superconductivity emerging near competing electronic orders [2], [3]. In these materials, superconductivity develops from a parent state characterized by itinerant antiferromagnetism in the form of a spin-density wave (SDW), arising from Fermi surface nesting. The proximity of magnetic and superconducting phases in their phase diagrams, together with their multiband electronic structure, points toward a pairing mechanism mediated by spin fluctuations. Consequently, the superconducting state is highly sensitive to disorder and impurity scattering [4], [5], [6], [7]. Among IBS families, the $RE$FeAsO (1111) compounds exhibit the highest transition temperatures, reaching ~58 K via fluorine substitution at the oxygen site. Their structural and chemical flexibility enables controlled tuning of carrier concentration and electronic correlations [8], [9]. Among these doping, the substitution in the superconducting FeAs layer is always quite interesting to understand the effect of replacement of Fe with other elements [10]. In particular, substitution at the Fe site within the superconducting FeAs layer provides a direct means to probe the interplay between magnetism and superconductivity. The impact of Fe-site substitution depends strongly on the nature of the dopant. Electron dopants such as Co, Ni, and Cu gradually suppress superconductivity while modifying the Fermi surface [11], [12], [13], [14]. Isovalent Ru primarily acts as a scattering center, leading to a relatively smooth reduction of $T_c$. In contrast, Mn substitution produces a markedly different response. In Ba(Fe$_{1-x}$Mn$_x$)$_2$As$_2$ and in La- and Sm-based 1111 systems, even very small Mn concentrations abruptly suppress superconductivity [15], [16], [17], [18], [19], [20], [21]. In LaFeAsO, Mn concentrations as low as $x\sim0.001–0.01$ fully destroy superconductivity, indicating that Mn behaves as a strong magnetic impurity rather than a weak potential scatterer.

Experimental studies show that Mn forms localized magnetic moments that strongly couple to itinerant Fe 3d electrons. In fluorine-doped LaFeAsO, Mn substitution restores stripe-type antiferromagnetic order and the associated tetragonal-to-orthorhombic structural transition, demonstrating that the system lies close to a magnetic instability. This behavior can be understood within an Ruderman-Kittel-Kasuya-Yosida (RKKY)-type framework, where Mn moments interact through exchange mediated by the enhanced spin susceptibility of the correlated FeAs planes. Moreover, transport studies in Mn-doped Ba-122 compounds reveal Kondo-like scattering at low concentrations, highlighting the role of local moment–itinerant electron interactions in reconstructing the electronic ground state. These effects sharply contrast with those of non-magnetic substitutions, underscoring the decisive influence of Mn-induced magnetism in destabilizing superconductivity [19], [20], [22], [23]. Despite extensive investigations in La- and Sm-based 1111 compounds, Mn substitution in fluorine-doped PrFeAsO remains largely unexplored. Positioned between La and Sm in the rare-earth series, Pr-based 1111 offers an opportunity to examine how Mn-induced magnetism competes with superconductivity in a system with intermediate electronic correlations [16], [20], [24].

In this work, we present a systematic investigation of the $PrFe_{1-x}Mn_xAsO_{0.7}F_{0.3}$ system, with Mn concentration $0 \leq x \leq 0.1$. By combining structural, Raman spectroscopy, transport, magnetotransport and magnetic measurements, we map the suppression of superconductivity and the evolution of impurity-induced magnetism. We observe a systematic decrease in the superconducting transition temperature from $T_c$ = 48.3 K in the parent compound to complete suppression at $x = 0.1$, accompanied by the development of a low-temperature resistivity upturn evolving into insulating-like behavior at high Mn content. Magnetization and magnetotransport measurements further reveal monotonic suppression of bulk superconductivity, critical current density $J_c$, and upper critical field $H_{c2}$ with increasing Mn concentration. Raman spectroscopy provides direct local evidence of Mn incorporation into the FeAs planes via progressive suppression of Fe-related phonon modes. Collectively, these results disentangle intrinsic pair-breaking effects from extrinsic microstructural contributions, offering a comprehensive understanding of how Mn-induced magnetic impurities perturb superconductivity across the rare-earth series.

## II. EXPERIMENTAL METHODS

Polycrystalline samples of $PrFe_{1-x}Mn_xAsO_{0.7}F_{0.3}$ were synthesized using a standard two-step solid state process (CSP) with nominal Manganese contents $x$ = 0, 0.01, 0.02, 0.03, 0.04, 0.05, 0.07 and 0.1. The starting precursors of high purity powders Pr(3N), $PrF_3$(3N), $Fe_2O_3$(4N), Fe(4N), As chunks(6N), and Mn(3N) powder were utilized and mixed thoroughly according to their nominal stoichiometry in an agate mortar and pestle for 10-15 mins and then thoroughly ground and pelletized under a uniaxial pressure ~200 bars. The synthesis was conducted in a high-purity argon glove box to avoid any contamination from oxygen and moisture. The obtained pellets were placed in a clamped tantalum (Ta) tube, which was sealed in an evacuated quartz tube under a vacuum of $10^{-2}$-$10^{-3}$ bar. In the first heating profile, the samples were heated at 520 °C for 8 h. The resulting pellets were reground, repressed and sealed again in an evacuated quartz tube under same conditions. In the second heating profile, the samples were heated at 950 °C for 12 h, followed by slow cooling at 110 °C over 18 h before the samples finally cooling down to the room temperature. The resulting samples were cut manually into rectangular bars for further structural, microstructural, transport, Raman and magnetic measurements.

For the structural analysis, powder X-ray diffraction (XRD) measurements were performed by employing a Panalytical Empyrean diffractometer with a step size of 0.013° within the 2$\theta$ range of 10° to 90° and a counting time of 300 s per step utilizing CuK$\alpha$ radiation ($\lambda$ = 1.5418 A) operated at 35 mA and 40 kV. The diffraction data was analyzed using PDF4+2025 database (ICDD) to evaluate the phase purity and extraction of diffraction profiles. Magnetic properties of the samples were measured using a vibrating sample magnetometer (VSM) in a physical property measurement system (PPMS, Quantum Design). The temperature dependent profiles of magnetic susceptibility were measured in zero-field-cooling (ZFC) and field-cooling (FC) modes under an applied magnetic field of 20 Oe across the temperature range of 5-60 K. Magnetic hysteresis (*M-H*) loops were recorded at 5 K in magnetic fields upto 9 T. For the electrical resistivity measurements in a zero-field, the measurements were carried out in a standard four-probe setup on rectangular pellets with typical dimensions of 2×3×2 mm$^3$ using a closed-cycle refrigerator or a PPMS across the temperature range of 7-300 K. Magnetotransport measurements were conducted for selected compositions in the temperature range 5-60 K in magnetic fields upto 9 T. The Raman scattering measurements were performed using a LabRam ARAMIS (Horiba Jobin Yvon) spectrometer. More details of the Raman spectroscopy and density functional theory (DFT) calculation are provided elsewhere [14], [25], [26]. Structure geometries and atomic forces were determined by the density functional theory (DFT) using the all-electron full-potential linearized augmented plane-wave (LAPW) program WIEN2k [27]. Perdew, Burke and Ernzerhof (PBE) form of the generalized gradient approximation (GGA) to the exchange-correlation functional [28] was utilized. Plane-wave cut-off parameter Rk$_{max}$ was set to 9, 2000 k-points in the first Brillouin zone were selected, and 6 nonspherical matrix elements for large spheres (LVNS) were taken. The non-default criterion for SCF convergences was charge difference less than 0.0001e. The rest of the initialization parametrs were set at default values. Lattice constants for PrFeAsO were taken from [29]. RMT spheres in the unit cell were reduced at 2% to: 2.38 for Pr, 2.21 for Fe, 2.10 for As, and 1.94 for O. Internal free z-parameters of Wyckoff 4c positions for Pr and As atoms were optimized by minimization of internal forces by WIEN2k MSR1a algorithm with SCF convergence criterion of force magnitude less than 0.1 mRy/a.u. Frequencies of Raman modes were then determined by the frozen-phonon method. Symmetric Raman-active modes were applied, determined by group theoretical method [30]. Force constants were obtained from the first derivative of 7-point linear interpolation of force vs. atom displacement relationship, for a series of displacements with ±0.01Å, ±0.02Å and ±0.03Å magnitudes.

## III.  RESULTS AND DISCUSSIONS

### a) Structural Analysis

Powder X-ray diffraction of $PrFe_{1-x}Mn_xAsO_{0.7}F_{0.3}$ samples with the nominal Mn content $x = 0$, 0.01, 0.02, 0.03, 0.04, 0.05, 0.07 and 0.1 are presented in Figure 1(a). All samples, including the parent compound ($x = 0$) and Mn-substituted compositions, predominantly crystallize in the tetragonal ZrCuSiAs-type structure with space group *P4/nmm*, which is characteristic of the 1111 family of IBS. For the parent compound ($x = 0$), secondary phases corresponding to PrOF (~9%), PrAs (~1%), and FeAs (~14%) are observed. In comparison, Mn-doped samples exhibit slightly improved phase purity and contain only minor impurity phases, primarily PrOF, PrAs, and MnO. The relative volume fractions of these impurities vary non-monotonically with Mn concentration. Quantitative analysis indicates that the impurity content ranges from approximately 1–9% for PrOF, 1–4% for PrAs, and 0–4% for MnO. Notably, the FeAs impurity phase disappears upon Mn substitution, highlighting the sensitivity of fluorine-doped 1111 compounds to synthesis conditions [31]. Moreover, no additional diffraction peaks associated with secondary FeAs-based superconducting phases are detected, indicating that Mn incorporation at the Fe sites does not destabilize the host crystal structure within the investigated doping range [1]. Overall, the total volume fraction of impurity phases remains significantly lower than that of the dominant Pr1111 phase, confirming that high structural integrity is preserved across the entire Mn-doping series.

A quantitative phase analysis was carried out for all compositions using Rietveld refinement implemented in the GSAS software package. The refined structural parameters were obtained with satisfactory reliability factors. A representative refinement profile for the $x = 0.01$ sample is presented in Figure 1(b), while refinement results for higher Mn contents ($x = 0.03$ and $0.05$) are shown in Supplementary Figure S1. In all cases, the experimental diffraction patterns are in good agreement with the calculated profiles, confirming the presence of only minor impurity phases, as discussed in relation to Figure 1(a). The refined lattice parameter $c$ and unit-cell volume $V$ for PrFe$_{1-x}$Mn$_x$AsO$_{0.7}$F$_{0.3}$ as a function of Mn concentration are displayed in Figures 1(c) and 1(d), respectively. For comparison, the corresponding data for Mn-doped Sm1111 samples are also included. Both the $c$-axis lattice parameter and the unit-cell volume $V$ exhibit a gradual increase with increasing Mn content relative to the parent compound ($x = 0$), which possesses lattice parameters $a = b = 3.978(6)$ Å and $c = 8.607(1)$ Å. This systematic expansion is consistent with the substitution of Mn at Fe sites within the FeAs layers, considering the larger ionic radius of Mn compared to Fe. Furthermore, the observed trends in '$c$' and '$V$' are in good agreement with previous reports on Mn-doped SmFe$_{1-x}$Mn$_x$AsO$_{0.88}$F$_{0.12}$ [12], [21], providing additional support for effective Mn incorporation into the Fe sublattice. The monotonic expansion along the $c$-axis and the corresponding increase in unit-cell volume $V$ indicate that lattice expansion is an intrinsic feature of the 1111 structure induced by larger dopant ions in the superconducting FeAs planes, despite the presence of minor impurity phases [1]. Energy-dispersive X-ray (EDX) spectroscopy was employed to examine the spatial distribution of Mn and assess compositional homogeneity. Representative elemental mapping images for samples with $x = 0$, 0.02 and 0.05 are shown in Supplementary Figure S2. The results reveal that Mn is nearly homogeneously distributed throughout the samples, with no evidence of Mn-rich regions or clustering, particularly at low substitution levels. This observation indicates that Mn preferentially occupies Fe sites within the FeAs planes rather than forming segregated impurity phases. With increasing Mn content, a gradual enhancement of microstructural inhomogeneity is observed, manifested by increased grain boundary roughness and local compositional fluctuations. Such microstructural features are expected to significantly influence the electrical transport properties of these polycrystalline samples through enhanced grain boundary scattering and reduced intergranular connectivity, rather than through substantial modifications of the intrinsic electronic band structure.

**b) Raman Spectroscopy**

Raman spectroscopy measurements were performed to probe local lattice dynamics and to investigate the effects of Mn substitution on the vibrational properties of PrFe$_{1-x}$Mn$_x$AsO$_{0.7}$F$_{0.3}$. This technique is particularly sensitive to changes in local bonding environments and atomic substitutions within specific sublattices. Figure 2(a) presents Raman spectrum of PrFe$_{1-x}$Mn$_x$AsO$_{0.7}$F$_{0.3}$ with the undoped sample ($x = 0$), revealing three distinct peaks located at 162 cm$^{-1}$, 204 cm$^{-1}$, 211 cm$^{-1}$. These modes are tentatively assigned to the out-of-plane lattice vibrations of Pr (A$_{1g}$), As (A$_{1g}$) and Fe (B$_{1g}$), respectively. To the best of our knowledge, no experimental or theoretical reports on the Raman spectrum of pristine PrFeAsO or doped

PrFeAs(O,F) are currently available in the literature that could serve as a reference for comparison with our measurements. To support the experimental interpretation, density functional theory (DFT) calculations were performed to estimate the phonon frequencies associated with collective atomic vibrations within the praseodymium, arsenic, and iron layers in pristine PrFeAsO (without fluorine doping). The calculated phonon frequencies show reasonably good agreement with the experimentally observed values, yielding 173 cm$^{-1}$ (+7%), 201 cm$^{-1}$ (-1.5%) and 202 cm$^{-1}$ (-4%), for Pr-, As- and Fe-related phone modes respectively. The small discrepancies between calculated and measured frequencies most likely originate from the presence of fluorine doping in the experimental samples, which may influence the lattice dynamics. Figure 2(b) presents further the evolution of the Raman-active phonon modes as a function of Mn concentration ($X_{Mn}$). A systematic softening of the Fe-related $B_{1g}$ mode is observed with increasing Mn content, characterized by a decrease in frequency of approximately 3 cm$^{-1}$ as $x$ increases from 0.01 to 0.10. This behaviour indicates a progressive perturbation of the Fe sublattice and provides strong evidence for the substitution of Mn at Fe sites within the FeAs planes. The observed phonon softening can be attributed to modifications of local bonding strength as well as mass effects associated with Mn incorporation. In contrast, the Pr-related $A_{1g}$ mode exhibits a non-monotonic evolution with Mn doping. A gradual decrease in frequency of approximately 3.5 cm$^{-1}$ is observed up to $x = 0.07$, followed by a sudden increase of about 2 cm$^{-1}$ at $x = 0.10$. This behaviour suggests that the PrO layers are indirectly influenced by Mn substitution at low and intermediate doping levels, while a more pronounced modification of the interlayer environment occurs at higher Mn concentrations. Such changes may be associated with variations in interlayer coupling and local strain effects induced by lattice expansion. The As-related $A_{1g}$ mode remains nearly unchanged throughout the investigated doping range, indicating that the Fe–As framework retains its structural integrity and is not significantly distorted by Mn substitution. This stability reflects the robustness of the FeAs tetrahedral network against moderate chemical perturbations. Overall, the pronounced sensitivity of the Fe-related modes, combined with the relative stability of the As modes and the non-monotonic evolution of the Pr-modes, provides compelling local-scale evidence for effective Mn incorporation within the FeAs planes. Furthermore, the observed modifications in the PrO-related vibrations are attributed primarily to changes in interlayer interactions rather than direct Mn substitution, highlighting the layered nature of the 1111 crystal structure. The modifications of lattice dynamics observed in this study for PrFe$_{1-x}$Mn$_x$AsO$_{0.7}$F$_{0.3}$, where the Fe layer is progressively substituted by manganese atoms, appear to contrast with the recent observations reported for SmFeAsO$_{0.8}$F$_{0.2}$ doped with the non-magnetic element, copper [14].

### c) Transport property

The temperature dependence of normal-state-resistivity at the room temperature for PrFe$_{1-x}$Mn$_x$AsO$_{0.7}$F$_{0.3}$ samples are shown in Figure 3(a). The resistivity of the parent sample $x = 0$ has room temperature resistivity ~5 mohm-cm, which decreases almost linearly before the superconducting transitions. A very small amount of Mn substitution, i.e. $x = 0.01$ has slightly higher resistivity than the parent sample $x = 0$ in the entire temperature regions, which could be due the extrinsic contribution such as grain connectivity and secondary phase such as the fraction of PrOF is relatively high, which can contribute to the insulating regions at the grain boundary. The highest resistivity is observed for $x = 0.01$, while the lowest resistivity occurs at $x = 0.03$ of the Mn content, followed by a more systematic increase for higher Mn concentrations, $x = 0.02, 0.04, 0.05, 0.07$ and $0.1$. Such trend in the resistivity suggests the role of multiple scattering mechanism governing the transport in polycrystalline samples; combined with the complexity of intrinsic scattering within the grain and extrinsic contributions from the grain boundaries and secondary phases [32]. By contrast, for $x = 0.03$ Mn content, the reduced resistivity suggests an improved effective grain connectivity, despite the impurities. A notable emergence of an upturn like anomaly is evident for the Mn substitutions $x \geq 0.04$ in the low temperature regime (below 100 K) upon cooling. The resistivity behaviour of the sample $x = 0.1$ are shown in the inset of figure 3(a) where the resistivity increases very rapidly below 100 K. Hence, upon further substitution of Mn, the observed upturn-like anomaly become more prominent leading to a metal-insulator-like behaviour for $x = 0.1$, and are commonly ascribed to the magnetic impurities, enhanced spin-fluctuations and likelihood localization driven by disorder. This low-temperature anomaly behaviour is comparable with other Mn substituted La-1111, Sm-1111 and Ba-122 iron-based superconductors [17], [19], [21]. Mn is known to instigate localized magnetic moment that couples with the electrons in the FeAs conduction layer, giving rise to the transport crossover from metallic to non-metallic at higher Mn substitutions [33]. The low temperature behaviour of these samples is shown in Figure 3(b). The superconducting transition of the parent sample $x = 0$ decreases systematically with Mn substitution. The superconducting transition is clearly observed for the sample $x = 0.07$. However, higher Mn substitution such as $x = 0$, no superconductivity is observed, as shown in the inset of Figure 3(b).

We have extracted the onset ($T_c^{onset}$) and offset ($T_c^{offset}$) superconducting transition temperatures, transition width ($\Delta T$), and Residual Resistivity Ratio ($RRR = \rho_{300K} / \rho_{60K}$) from the temperature-dependent resistivity measurements, as depicted in Figures 3(a)-(b). The superconducting transition temperature are determined using the tangent method applied to the resistivity curves. Specifically, $T_c^{onset}$ is obtained from the intersection between the extrapolated normal-state resistivity and the extrapolated steepest slope of the resistive transition, whereas $T_c^{offset}$ corresponds to the temperature at which zero resistivity is achieved. The variation of $T_c^{onset}$ as a function of nominal Mn content ($x$) is presented in Figure 3(c). A monotonic decrease in $T_c^{onset}$ is observed with increasing Mn concentration, decreasing from 48.3 K for $x = 0$ to 45.9 K, 38.5 K, 34.2 K, 27.2 K, 21.9 K, and 19.0 K for $x = 0.01, 0.02, 0.03, 0.04, 0.05$, and 0.07, respectively. For $x = 0.10$, superconductivity is completely suppressed within the measured temperature range. The rapid suppression of the superconducting transition temperature with Mn substitution is indicative of strong pair-breaking effects arising from magnetic impurity scattering within the FeAs planes [19], [20]. Manganese ions introduce localized magnetic moments, which strongly perturb the Cooper pairs and reduce superconducting coherence. Such behavior is consistent with previous reports on Mn-substituted 1111- and 122-type iron-based superconductors, confirming that Mn acts as an efficient magnetic pair breaker in these systems. Figure 3(d) illustrates the variation of the superconducting transition width, $\Delta T$, defined as ($T_c^{onset} - T_c^{offset}$), as a function of Mn concentration ($x$). For low Mn doping levels ($x = 0.01$ and 0.02), $\Delta T$ increases approximately linearly, rising from 6.5 K for the parent compound ($x = 0$) to about 8.5 K for $x = 0.02$. With further Mn substitution ($x = 0.03, 0.04, 0.05$, and 0.07), a slight reduction in $\Delta T$ is observed; however, the transition width remains significantly broader than that of the parent sample. Overall, the dependence of $\Delta T$ on Mn content exhibits a dome-like behavior, with a maximum in the intermediate doping range ($x \approx 0.02–0.04$). The systematic broadening of the superconducting transition in Mn-substituted samples reflects enhanced spatial inhomogeneity and increased disorder arising from Mn-induced magnetic correlations and impurity scattering. These effects lead to variations in local superconducting properties and promote percolative transport behavior. The modest reduction in $\Delta T$ at higher Mn concentrations suggests a competition between increasing impurity scattering and the progressive suppression of superconductivity. As the superconducting volume fraction diminishes with increasing Mn content, percolative effects become less pronounced, resulting in a partial narrowing of the resistive transition despite the presence of substantial disorder. The variation of the residual resistivity ratio (*RRR*) as a function of Mn concentration is shown in Figure 3(e). *RRR* decreases rapidly for low Mn doping levels ($x \leq 0.05$) and becomes nearly constant for higher Mn contents, approaching a value close to 1 for $x = 0.10$. This behavior indicates a pronounced degradation of metallic transport upon Mn substitution, reflecting the dominance of intrinsic impurity scattering within the FeAs planes. The systematic reduction of *RRR* is accompanied by the emergence of a broad resistivity upturn at low temperatures for higher Mn concentrations, as illustrated in Figure 3(a). Such behavior contrasts with the effects of other magnetic impurities, such as Co, Ni, Rh, or Pd, at Fe sites, where transport anomalies are typically associated with disorder-induced weak localization [34]. In the present case, the strong magnetic scattering by localized Mn moments interacting with itinerant Fe-3d electrons [19] appears to play a central role. The observed transport trends suggest a possible redistribution of spectral weight near the Fermi level and a suppression of interband scattering mechanisms linked to Fermi surface nesting [33]. The Mn-induced perturbations of the nested hole and electron pockets are likely to weaken spin-fluctuation-mediated pairing, while promoting incoherent charge transport. Consequently, the low-temperature resistivity upturn observed for $x \geq 0.04$ and the insulating-like behavior for $x = 0.10$ reflect a progressive suppression of electronic states near the Fermi energy, consistent with reports on Mn-substituted Ba-122 systems. Although direct probes of the Fermi surface are beyond the scope of this work, the transport results are internally consistent with Mn-induced modifications of the nested electronic structure mediated by magnetic correlations. The combined Raman and resistivity data provide a coherent picture of Mn acting as a strong localized perturbation within the FeAs planes. The progressive softening of Fe-related phonon modes with increasing Mn content indicates a degradation of metallicity and suppression of superconductivity, demonstrating that both lattice dynamics and electronic degrees of freedom are significantly affected by Mn substitution at Fe sites. Furthermore, the non-monotonic evolution of the Pr-related phonon modes, with a sudden increase in frequency at higher Mn concentrations, suggests a crossover in the coupling between the PrO layers and the FeAs planes. This interpretation is corroborated by the low-temperature resistivity upturns observed in samples with high Mn content, highlighting the interplay between lattice, electronic, and magnetic perturbations induced by Mn doping.

Magnetotransport measurements were performed to investigate the influence of Mn substitution on vortex dynamics and superconducting properties in $PrFe_{1-x}Mn_xAsO_{0.7}F_{0.3}$ samples. Figures 4(a) and 4(b) present the resistive superconducting transitions under applied magnetic fields up to 9 T for $x$ = 0.01 and 0.03, respectively. Corresponding data for the parent compound ($x$ = 0) have been reported previously [35]. For both parent and Mn-substituted samples, the superconducting transition shifts systematically toward lower temperatures and exhibits progressive broadening with increasing magnetic field, reflecting enhanced vortex dissipation and flux-flow resistivity. To determine the upper critical field ($H_{c2}$) and irreversibility field ($H_{irr}$), a consistent resistive criterion based on 90% and 10% of the normal-state resistivity at $T = T_c$ was adopted, respectively [8]. The resulting temperature derivatives of $H_{c2}$ and $H_{irr}$ are summarized in Figure 4(c) for $x$ = 0, 0.01, and 0.03. With Mn substitution, the slope of the upper critical field near $T_c$: $\frac{dH_{c2}}{dT}|_{45\ K}$ and $\frac{dH_{c2}}{dT}|_{34\ K}$ decreases from $\frac{dH_{c2}}{dT}|_{48\ K}$ = −6.30 T/K for the parent compound ($x$ = 0) to −8.22 T/K and −6.32 T/K for $x$ = 0.01 and 0.03, respectively. Similarly, the slope of the irreversibility field, ($\frac{dH_{irr}}{dT}$), is reduced to −1.58 T/K and −1.91 T/K for $x$ = 0.01 and 0.03, respectively, compared to −1.30 T/K for $x$ = 0. The upper critical field $H_{c2}$ of PrFe$_{1-x}$Mn$_x$AsO$_{0.7}$F$_{0.3}$ samples was estimated within the framework of the single-band Werthamer–Helfand–Hohenberg (WHH) model [36]. The extrapolated orbital-limiting field yields $H_{c2}(0)$ ~261 T for $x$ = 0.01 and ~150 T for $x$ = 0.03, comparable to values reported for other 1111- and 122-type iron-based superconductors [12], [37] characterized by enhanced impurity scattering. The non-monotonic evolution of $H_{c2}$ with Mn content reflects a competition between disorder-induced scattering and magnetic pair-breaking effects. Using the Ginzburg-Landau relation for coherence length $\xi_{GL}(0) = \sqrt{\frac{\Phi_0}{2\pi\mu_o H_{c2}(0)}}$ [38]; the corresponding zero-temperature coherence length are estimated to be approximately 1.1 nm and 1.5 nm for $x$ = 0.01 and 0.03, respectively. These values are significantly smaller than those expected for clean-limit BCS superconductors [37], indicating that the Mn-substituted Pr-1111 system reside in the dirty limit, similar to the parent system ($x$ = 0) [35]. Furthermore, the mean free path, estimated from $\xi \sim \sqrt{\xi_0 l}$ [38], is found to be in the range of $l$ = 0.2-0.3 nm, comparable to the in-plane Fe–Fe distance ($d_{Fe-Fe} = \frac{a}{\sqrt{2}} = 0.28\ nm$), where $a$-lattice parameter (~ 4 Å) is derived from the structural data exhibiting tetragonal crystal structure. This places the system close to the Ioffe–Regel limit [39], establishing Mn as a strong scattering center, when substituted at Fe sites. Such strong impurity scattering enhances the initial slope $\frac{dH_{c2}}{dT}|_{T=T_c}$ and increase $H_{c2}$ at low Mn concentration ($x$ = 0.01), consistent with multiband superconductivity. However, at higher Mn concentration ($x$ = 0.03), magnetic pair-breaking effects become dominant. Unlike Co or Ni substitutions, Mn carries localized magnetic moments, leading to strong spin-flip scattering and rapid suppression of superconductivity [12]. The concomitant reduction of $T_c$ and $H_{c2}$ at $x = 0.03$ reflects competition between superconductivity and short-range magnetic correlations within the FeAs planes. The Pauli-limiting field, estimated using $H_p = 1.84 \cdot T_c$, yields values of approximately 84.5 T for $x = 0.01$ and 63 T for $x = 0.03$. The corresponding Maki parameters, $\alpha\ (= \sqrt{2}\,\frac{H_c^{orb}(0)}{H_P})$ [40], are estimated to be about ~4.4 for $x = 0.01$ and ~3.3 for $x = 0.03$, respectively, indicating a strong Pauli paramagnetic regime ($\alpha > 1$). The significantly larger orbital critical fields compared to the Pauli limit highlight the important roles of multiband effects and enhanced spin–orbit scattering in stabilizing superconductivity against paramagnetic pair breaking, consistent with behavior observed in the parent compound $x = 0$.

Figure 5 presents the analysis of thermally activated flux flow (TAFF) behavior associated with resistive broadening under applied magnetic fields. Within the TAFF framework, the resistivity in the mixed state follows an Arrhenius-type relation: $\rho = \rho_0\,exp\left(-\frac{U_0(T,B)}{k_B T}\right)$; where $U_0$ is the effective activation energy for vortex motion and $k_B$ is the Boltzmann constant [41]. Figure 5(a) and (b) represents Arrhenius plots of ln ($\rho/\rho_0$) vs $1/T$ for $x = 0.01$ and 0.03 respectively, under the magnetic fields up to 9 T. Here, $\rho_0$ represents the normal-state resistivity at 60 K and zero field. In both cases, extended linear regions are observed over a broad temperature range, validating the applicability of the TAFF model. The effective activation energy $U_0$ was extracted from the slopes of these linear regions and analyzed as a function of magnetic field. The field dependence of the activation energy follows a power-law relation, $U_0 \sim H^{-\eta}$. For the parent compound ($x = 0$), two distinct regimes are identified, characterized by exponents $\eta_1 = 0.3$ at low fields and $\eta_2 = 0.69$ at high fields, as shown in Figure 5(c). This crossover behavior is commonly observed in iron-based superconductors and is typically associated with a transition from single-vortex pinning at low fields to weak collective pinning at higher fields [37]. Upon Mn substitution, the field dependence of $U_0$ is significantly weakened. For $x = 0.01$, the exponents decrease to $\eta_1 = 0.24$ and $\eta_2 = 0.5$, while further suppression is observed for $x = 0.03$, with $\eta_1 = 0.16$ and $\eta_2 = 0.32$ in the low- and high-field regimes, respectively. The systematic reduction of both $\eta_1$ and $\eta_2$ with increasing Mn content indicates a gradual evolution toward a more dissipative vortex regime characterized by enhanced vortex mobility and weakened pinning efficiency. Notably, the magnitude of the effective activation energy $U_0$ for $x = 0.01$ is slightly higher than that of the parent and $x = 0.03$ samples, as illustrated in Figure 5(c). This enhancement should be interpreted as an improvement in the effective energy barrier for vortex motion rather than an indication of strengthened intrinsic superconductivity. In polycrystalline superconductors, such as $YBa_2Cu_3O_7$ [42] or FeSe [43], TAFF behavior is known to be strongly influenced by intergranular coupling, intragrain pinning, and microstructural features. For low Mn substitution ($x = 0.01$), subtle modifications in microstructure and grain connectivity may enhance the effective vortex pinning landscape, leading to a transient increase in $U_0$. With further Mn incorporation ($x = 0.03$), disorder and magnetic pair-breaking effects become dominant, resulting in reduced activation energy and weaker field dependence, consistent with enhanced vortex dissipation. The observed evolution of the exponent $\eta$ with Mn content is comparable to that reported for other fluorine-doped 1111 systems with varying disorder levels [37], corroborating that Mn-induced magnetic correlations weaken collective pinning and promote dissipative vortex dynamics. Overall, the TAFF analysis demonstrates that Mn substitution systematically modifies both the magnitude and field dependence of the vortex activation energy, revealing a progressive crossover from relatively strong pinning in the parent compound to increasingly dissipative vortex motion at higher Mn concentrations.

### d) Magnetization and critical current density

Magnetic measurements were performed in the temperature range of 5–60 K under a constant applied magnetic field of 20 Oe using zero-field-cooled (ZFC) and field-cooled (FC) protocols to investigate superconducting properties and intergranular coupling in $PrFe_{1-x}Mn_xAsO_{0.7}F_{0.3}$ samples. Figure 6(a) displays the temperature dependence of the normalized magnetic moment, $M/M_{5K}$, for Mn concentrations $x = 0, 0.01, 0.02, 0.03, 0.04$ and $0.5$. The magnetization data reveal a systematic suppression of bulk superconductivity with increasing Mn content. The magnetic transition temperature, $T_c$, was determined from the onset of bifurcation between the ZFC and FC branches, providing a reliable estimate of the bulk superconducting transition. The extracted $T_c$ decreases monotonically from 46.1 K for $x = 0$ to 44.8 K, 37.5 K, 33.5 K, 25.9 K, and 19.8 K for x = 0.01, 0.02, 0.03, 0.04, and 0.05, respectively. These values are approximately 1–2 K lower than those obtained from resistivity measurements, consistent with previous reports on iron-based superconductors [31]. The pronounced and monotonic suppression of superconductivity confirms that Mn acts as an efficient magnetic pair-breaking impurity in the $PrFe_{1-x}Mn_xAsO_{0.7}F_{0.3}$ system. In contrast, non-magnetic substitutions at Fe sites, such as isovalent Ru doping in $SmFe_{1-x}Ru_xAsO_{0.85}F_{0.15}$ [44], exhibit significantly weaker effects on superconductivity. This comparison highlights the dominant role of magnetic scattering in destabilizing the superconducting state within the FeAs planes of Mn-substituted samples. For the parent compound ($x = 0$), a pronounced separation between the ZFC and FC branches is observed below $T_c$, a feature commonly reported in polycrystalline 1111 materials and typically attributed to weak-link behavior [31], [45]. This behavior arises from weak intergranular coupling, flux trapping, and inhomogeneous current pathways associated with grain boundaries and secondary phases, as also indicated by the transport measurements. Upon Mn substitution, the ZFC–FC curves become progressively smoother and exhibit a more single-step transition, despite the overall reduction in $T_c$. This qualitative improvement suggests enhanced intergranular coupling and reduced weak-link effects in Mn-doped samples. Such behavior is consistent with microstructural and compositional analyses, which reveal a nearly homogeneous distribution of Mn within the grains (Supplementary Figure S2). The improved grain connectivity likely mitigates weak-link effects and promotes more uniform superconducting screening currents, even though magnetic pair-breaking continues to suppress the intrinsic superconducting properties. Consequently, Mn substitution introduces a complex interplay between enhanced intergranular homogeneity and intrinsic magnetic scattering, governing the overall superconducting response of the system.

Magnetic hysteresis (*M-H*) measurements were performed at 5 K for selected compositions ($x = 0$, 0.01, and 0.02) to evaluate the critical current density properties. The hysteresis loop width, $\Delta m$, was determined from the difference in magnetic moments between the increasing and decreasing field branches in the first and second quadrants of the *M-H* loop. The critical current density $J_c$ is estimated using the Bean critical-state model, which relates $J_c$ to the width of the magnetic hysteresis loop. For rectangular samples, $J_c$ was calculated using the expression: $J_c = \frac{20 \Delta m}{Va\left(1-\frac{a}{3b}\right)}$ where *a* and *b* represent the sample dimensions ($a < b$), and *V* denotes the sample volume. Figure 6(b) shows a systematic reduction of $J_c$ with increasing Mn content, indicating a progressive degradation of current-carrying capability. This trend is consistent with the reduced superconducting volume fraction observed in other Mn-substituted 1111 systems [21], [46] and reflects the weakening of superconductivity due to magnetic pair-breaking effects. The presence of localized Mn moments disrupts Cooper pairing and weakens vortex pinning, thereby limiting the sustainable supercurrent density. The simultaneous observation of reduced $J_c$ and smoother magnetic transitions suggests that, although Mn substitution improves grain connectivity and suppresses weak-link behavior, it does not compensate for the intrinsic suppression of superconductivity. Consequently, the dominant factor governing the deterioration of current-carrying performance in PrFe$_{1-x}$Mn$_x$AsO$_{0.7}$F$_{0.3}$ remains the strong magnetic scattering introduced by Mn at the Fe sites, which overrides microstructural improvements and constrains the overall superconducting performance.

**e) Discussion**

Figure 7(a) presents a comparative analysis of the suppression of the superconducting transition temperature $T_c$ as a function of Mn concentration for LaFe$_{1-x}$Mn$_x$AsO$_{0.89}$F$_{0.11}$ (La1111) [19], SmFe$_{1-x}$Mn$_x$AsO$_{0.88}$F$_{0.12}$ (Sm1111) [12], NdFe$_{1-x}$Mn$_x$AsO$_{0.89}$F$_{0.11}$ (Nd1111) [47] and PrFe$_{1-x}$Mn$_x$AsO$_{0.7}$F$_{0.3}$ (Pr1111) (present work). Although superconductivity is suppressed in all 1111 systems upon Mn substitution, the rate of suppression varies substantially. Among these compounds, La-1111 exhibits the most rapid suppression, with superconductivity being destroyed at very low Mn concentrations, followed by Nd-1111 and Sm-1111. In contrast, the Pr-1111 system displays a comparatively slower suppression rate, with superconductivity persisting up to higher Mn contents, indicating enhanced robustness against Mn-induced perturbations. These systematic differences highlight the crucial role of the host electronic environment in governing the effectiveness of magnetic pair-breaking mechanisms within the 1111 family. In La-1111, Mn substitution is known to restore the magnetic ground state of the parent compound, leading to the re-emergence of stripe-type antiferromagnetic spin-density-wave fluctuations and rapid destruction of superconductivity [20]. This behavior reflects the proximity of the La-based system to magnetic instability. In comparison, Sm-1111 and Pr-1111 appear to accommodate Mn-induced carrier localization and disorder over a wider doping range before strong magnetic pair-breaking becomes dominant. Consequently, Mn acts not merely as a pair-breaking impurity, but also as a sensitive probe of the underlying electronic correlations and magnetic tendencies of the host material. Further comparison with MgB$_2$ and BaFe$_2$As$_2$-based systems [18], [17], [16], [48] underscores the exceptional effectiveness of Mn as a magnetic impurity in iron-based superconductors. Moreover, the systematic variation in rare-earth ionic size from La to Sm to Pr modulates the coupling between the FeAs planes and the rare-earth–oxide layers, thereby influencing the stability of superconductivity. This structural and electronic tuning plays a central role in determining the resilience of superconductivity against magnetic perturbations in the 1111 family.

To further elucidate the influence of Mn on normal-state transport, the evolution of the residual resistivity $\rho_0$, extracted at 60 K, is shown in Figure 7(b). A systematic increase in $\rho_0$ is observed with increasing Mn content, with a pronounced enhancement at higher doping levels. For example, at $x = 0.1$, $\rho_0$ reaches approximately 19 mΩ-cm. This data point is omitted from the main plot to emphasize the behavior at lower Mn concentrations. While moderate scattering is already present at low Mn levels, likely arising from microstructural variations and minor secondary phases, the substantial increase in $\rho_0$ at higher Mn content is consistent with the emergence of low-temperature resistivity upturns and insulating-like behavior. The close correlation between the increase in residual resistivity and the suppression of $T_c$ provides strong evidence that Mn acts as an efficient magnetic pair breaker in $PrFe_{1-x}Mn_xAsO_{0.7}F_{0.3}$. The observed suppression of superconductivity is consistent with impurity scattering effects described within the Abrikosov–Gor'kov framework (AG) for magnetic impurities [49], [50], as demonstrated by the fitting presented in the supplementary Figure S3. The simultaneous enhancement of normal-state anomalies and suppression of superconductivity further supports the central role of impurity-induced magnetic scattering. Overall, these results establish Mn substitution as an effective means of probing the interplay between magnetism, disorder, and superconductivity in fluorine-doped PrFeAsO. The systematic evolution of structural, transport, magnetic, and vortex-dynamical properties reveals that Mn-induced correlations and localized magnetic moments dominate the electronic response, providing valuable insight into the mechanisms governing superconductivity in the 1111 family.

## IV. CONCLUSION

We systematically investigated the effects of Mn substitution on the superconducting properties of $PrFe_{1-x}Mn_xAsO_{0.7}F_{0.3}$ over a wide doping range using complementary structural, Raman spectroscopic, transport, and magnetic measurements. Structural analysis confirmed that Mn predominantly substitutes at Fe sites, leading to a systematic expansion along the $c$-axis and an increase in unit-cell volume $V$ relative to the parent compound. These findings are corroborated by Raman spectroscopy, which reveals selective softening of Fe-related ($B_{1g}$) phonon modes while the As-related As-modes ($A_{1g}$) modes remain largely unaffected, confirming effective Mn incorporation within the FeAs planes. The normal-state transport evolves from metallic behavior in the parent compound to progressively incoherent and insulating-like characteristics at higher Mn concentrations, manifested by low-temperature resistivity upturns, reduced residual resistivity ratios, and suppressed superconducting transition temperatures. These results provide compelling evidence that Mn-induced magnetic correlations strongly interact with itinerant Fe-3d electrons, giving rise to pronounced magnetic pair-breaking and a gradual destruction of superconductivity. Magnetic measurements further demonstrated a monotonic reduction in bulk superconductivity, $J_c$, $H_{c2}$, and vortex activation energy with increasing Mn content, despite the observed improvement in intergranular coupling. This highlights the dominant role of magnetic impurity scattering in destabilizing the superconducting state, overriding microstructural improvements. A comparative analysis within the 1111 family reveals that the suppression of superconductivity in Pr-1111 is weaker than in Sm-1111 and significantly weaker than in La-1111, where superconductivity is rapidly destroyed at low Mn concentrations. This systematic trend reflects variations in Mn-induced magnetic correlations and rare-earth–FeAs plane coupling across the rare-earth series, which modulate the effectiveness of magnetic pair-breaking. Consequently, Pr-1111 emerges as a relatively robust host against Mn-induced perturbations, providing valuable insight into the complex interplay between magnetism, disorder, and superconductivity in IBS. Overall, this study establishes Mn substitution as a powerful probe of electronic correlations and magnetic interactions in fluorine-doped PrFeAsO and contributes to a deeper understanding of impurity effects in unconventional superconductors.

## CRediT authorship contribution statement

**Priya Singh:** Writing – review & editing, Writing – original draft, Investigation, Formal analysis, Data curation. **Konrad Kwatek:** Formal analysis, Investigation, Resources, Data curation, Writing – review & editing. **Tatiana Zajarniuk:** Data curation, Investigation, Resources, Writing – review & editing. **Taras Palasyuk:** Writing – review & editing, Formal analysis, Investigation, Data curation. **Cezariusz Jastrzębski:** Writing – review & editing, Resources, Data curation. **Andrzej Szewczyk:** Writing–review & editing, Resources, Data curation. **Michał Wierzbicki:** Writing–review & editing, Investigation, Resources, Methodology. **Shiv J. Singh:** Writing–review & editing, Writing – original draft, Visualization, Validation, Supervision, Software, Resources, Methodology, Investigation, Funding acquisition, Formal analysis, Conceptualization.

## Declaration of competing interest

The authors declare that they have no known competing financial interests or personal relationships that could have appeared to influence the work reported in this paper.

## Data availability

The raw/processed data required to reproduce these findings cannot be shared at this time due to technical or time limitations. Data are available upon request to the corresponding author.

**Acknowledgments:**

This work was funded by SONATA-BIS 11 project (Registration number: 2021/42/E/ST5/00262) and the Weave-UNISONO project (2025/07/Y/ST5/00116) sponsored by National Science Centre (NCN), Poland. SJS acknowledges financial support from National Science Centre (NCN), Poland through research Project number: 2021/42/E/ST5/00262 and 2025/07/Y/ST5/00116.

**Figure 1:** (**a**) Powder X-ray diffraction patterns of $PrFe_{1-x}Mn_xAsO_{0.7}F_{0.3}$ samples with nominal manganese Mn content for $x$ = 0, 0.01, 0.02, 0.03, 0.04, 0.05, 0.07 and 0.1 are shown. (**b**) Rietveld refinement of the room temperature XRD pattern for the sample $x$ = 0.01 are presented, depicting the experimental data, observed, calculated profile with the PrFeAsO (Pr1111) phase pattern, and the difference curve. The variation of (**c**) lattice parameter '$c$', and (**d**) lattice volume '$V$' as a function of the nominal Mn content ($x$) are shown. The possible error bars in the lattice parameters and the volume are included in panels (**c**)–(**d**). For comparison, the lattice parameters and unit cell volumes from Singh *et al.* [12] are also included in panels (**c**)–(**d**).

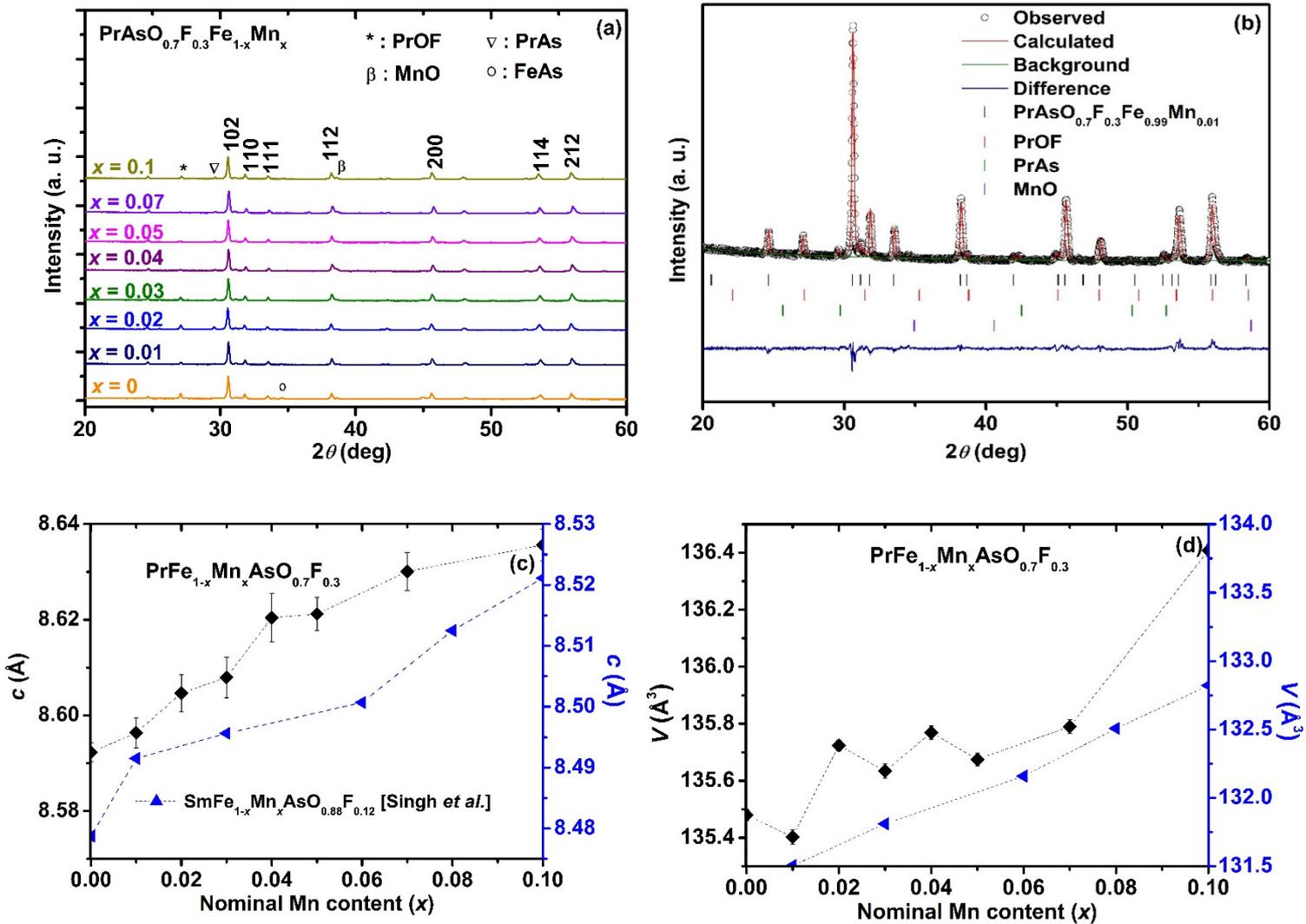

**Figure 2:** Raman scattering study of Mn-doped PrFe$_{1-x}$Mn$_x$AsO$_{0.7}$F$_{0.3}$ series. **(a)** Representative Raman spectra of the pristine compound ($x = 0$) and doped materials with manganese content 7% ($x = 0.07$) and 10% ($x = 0.1$). Spectra are vertically offset for clarity. Assignment of detected signals related to lattice vibrations is shown for the spectrum of the pristine compound. Deconvolution of experimental spectra with Lorentz function is shown as green lines. Total fits of Lorentz model to experimental data are shown by red lines. Vertical black lines are a guide to the eye. **(b)** Evolution of peak positions as a function of manganese content. Experimental data are shown as circles with corresponding error bars. Black solid lines are a guide to the eye.

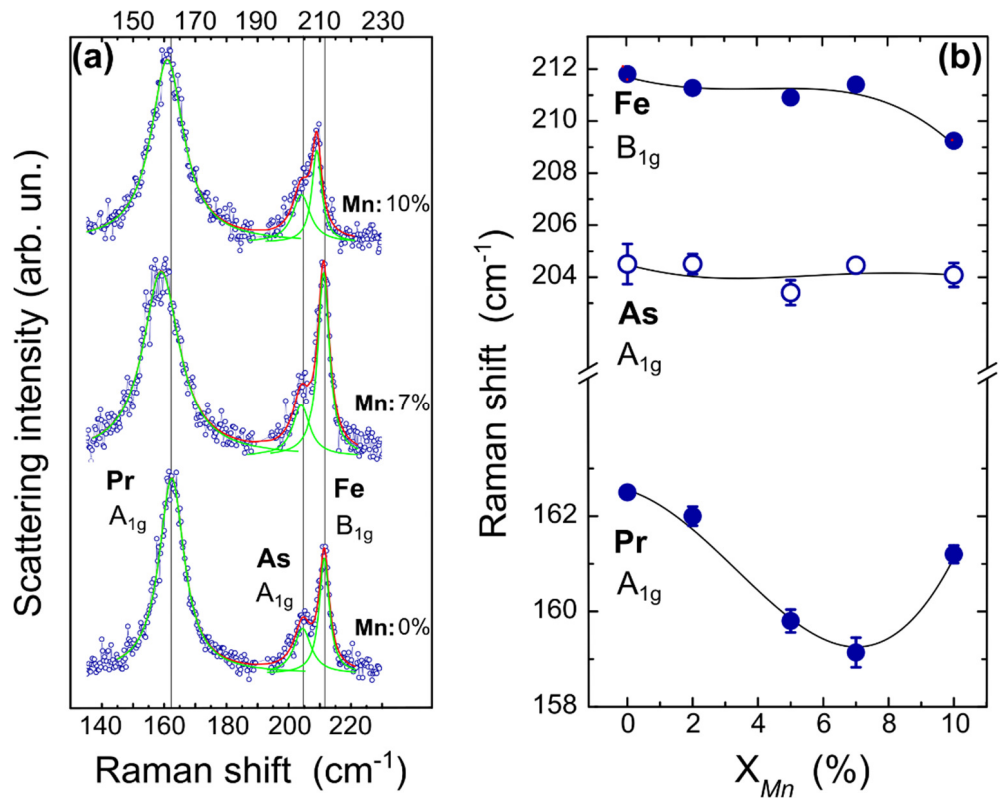

**Figure 3:** (**a**) The temperature dependence of resistivity $\rho(T)$ for $PrFe_{1-x}Mn_xAsO_{0.7}F_{0.3}$ samples for nominal Mn content $x$ = 0, 0.01, 0.02, 0.03, 0.04, 0.05, 0.07 in the temperature range 7-300 K. The inset figure of Figure (a) is plotted for $x$ = 0.1. (**b**) Expanded view of low temperature regime for all the Mn-containing compositions in the temperature range 7-50 K, with the inset figure representative of $x$ = 0.1. The variation of (**c**) onset of superconducting transition temperature $T_c^{onset}$, (**d**) transition width $\Delta T$ and (**e**) residual resistivity ratio (*RRR*) defined as $\rho_{300K}/\rho_{50K}$ is presented as a function of nominal Mn content ($x$).

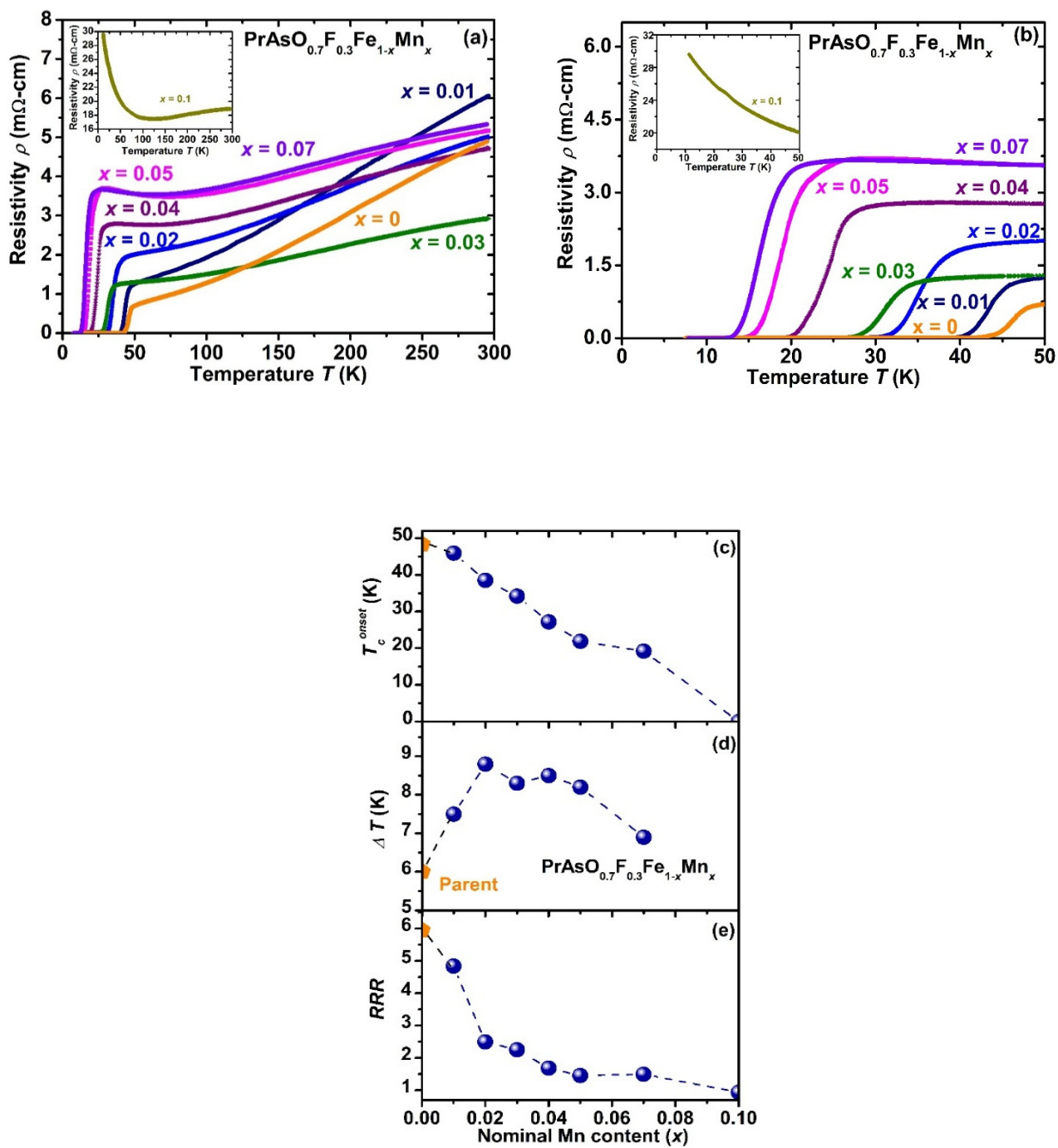

**Figure 4:** The temperature dependence of normalized resistivity $\rho/\rho_{50\,K}$ for PrFe$_{1-x}$Mn$_x$AsO$_{0.7}$F$_{0.3}$ samples under various applied magnetic fields upto 9 T is plotted for (**a**) $x = 0.01$, (**b**) $x = 0.03$ respectively. (**c**) The *H-T* phase diagrams showing upper critical field $H_{c2}$ and irreversibility field $H_{irr}$ as a function of temperature using a resistive criterion of 90% and 10% respectively as described in the text, is illustrated for $x = 0$, 0.01 and 0.03.

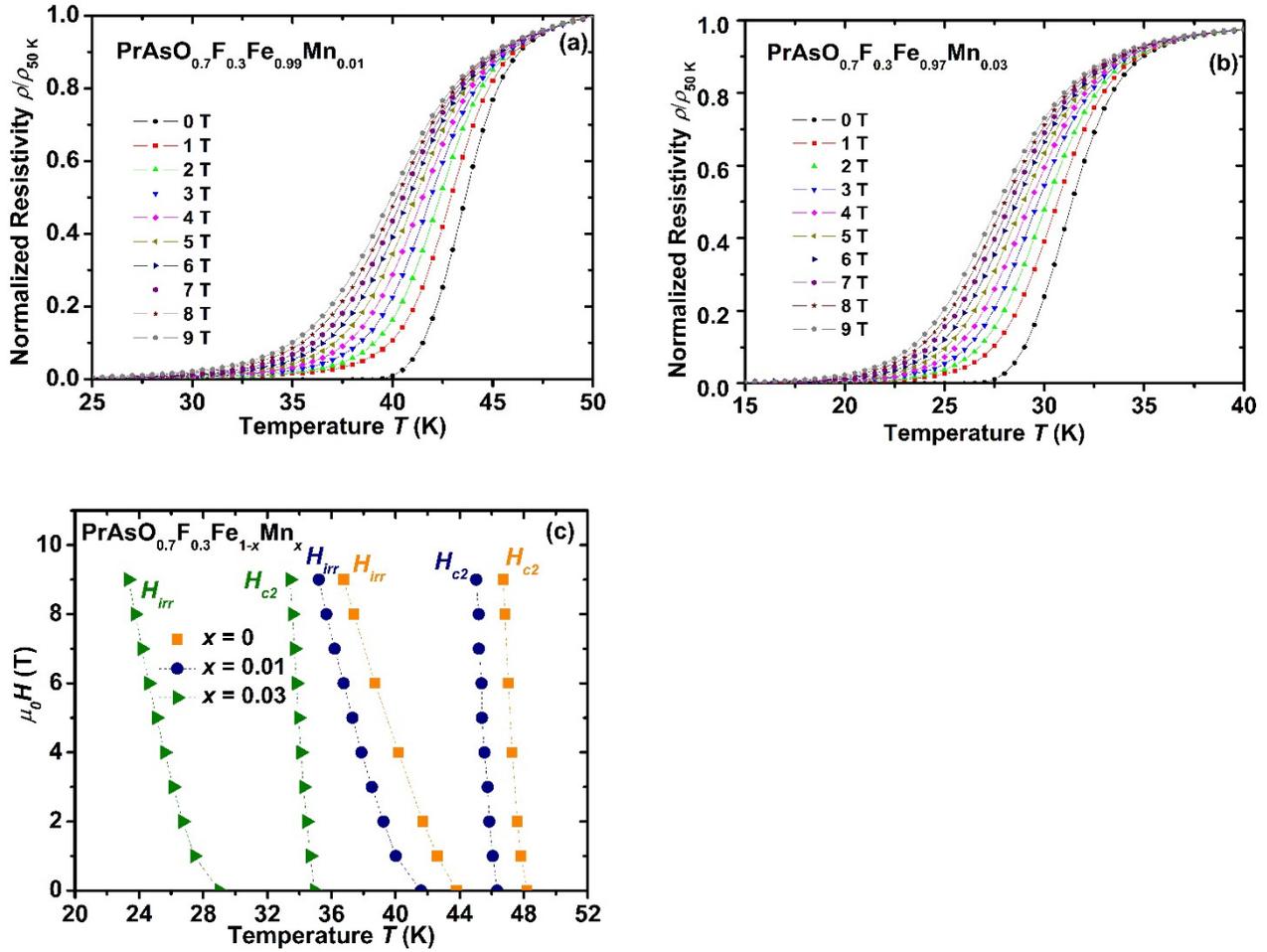

**Figure 5:** The Arrhenius plot of the normalized resistivity $log\,(\rho/\rho_0)$ as a function of inverse temperature $1/T$ under various applied magnetic fields up to 9 T for PrFe$_{1-x}$Mn$_x$AsO$_{0.7}$F$_{0.3}$ samples is illustrated for (**a**) $x = 0.01$ and (**b**) $x = 0.03$, respectively. Here $\rho_0$ is considered as the resistivity value at 60 K. (**c**) The field dependence of the extracted activation energy $U_0(H)$ for $x = 0$, 0.01 and 0.03, obtained from the slopes of the Arrhenius fit.

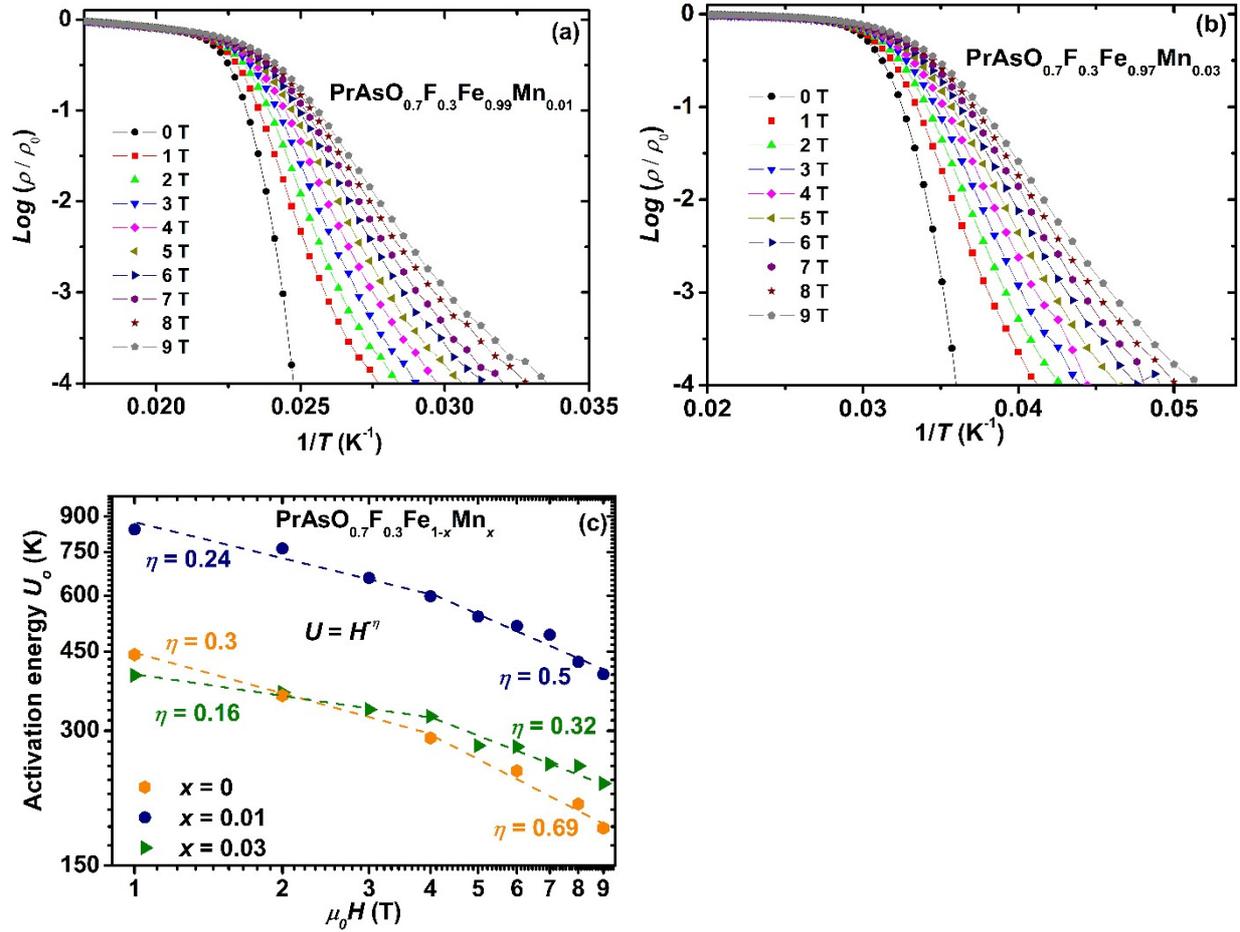

**Figure 6:** (**a**) The temperature dependence of the normalized magnetic moment $M/M_{5K}$ of zero-field-cooled (ZFC) and field-cooled (FC) modes for PrFe$_{1-x}$Mn$_x$AsO$_{0.7}$F$_{0.3}$ samples with the nominal Mn content $x$ = 0, 0.01, 0.02, 0.03, 0.04 and 0.05 recorded at magnetic field $H$ = 20 Oe. (**b**) The magnetic field dependence of critical current density $J_c$ is illustrated for $x$ = 0, 0.01 and 0.02 at 5 K.

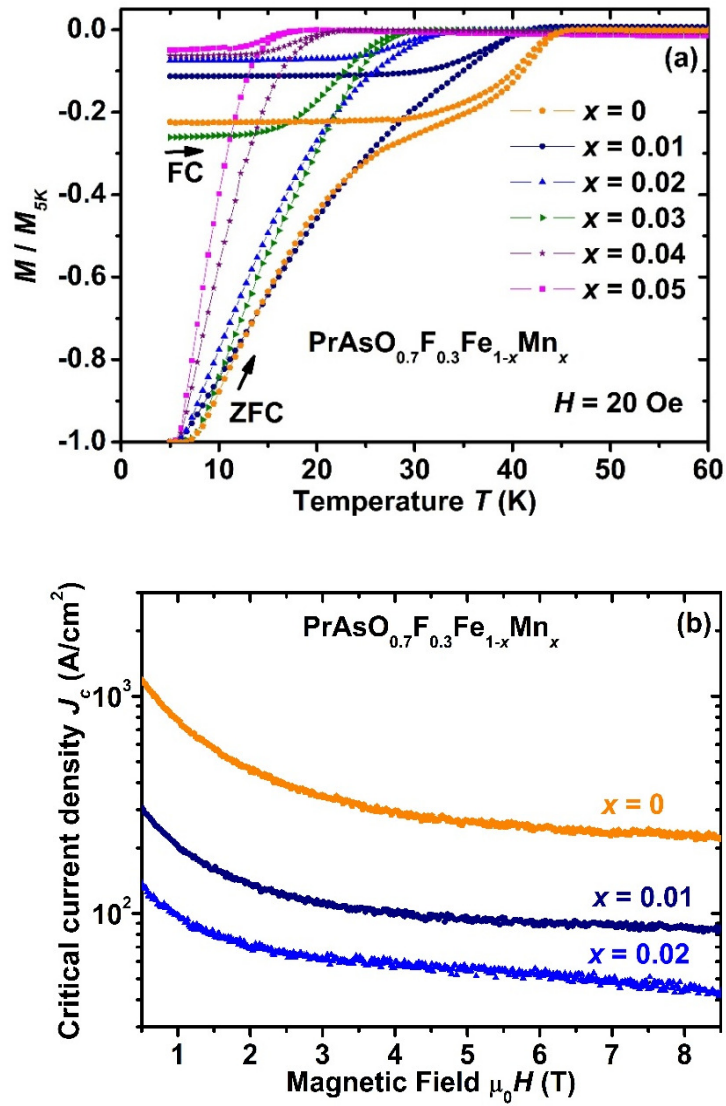

**Figure 7:** (**a**) The comparison of superconducting transition temperature $T_c$ as a function of Mn content ($x$) for the Mn substituted $RE$FeAs$_{1-x}$F$_x$ (1111) systems is illustrated for LaFe$_{1-x}$Mn$_x$AsO$_{0.89}$F$_{0.11}$ (Hammerath *et al.* [19]), SmFe$_{1-x}$Mn$_x$AsO$_{0.88}$F$_{0.12}$ (Singh *et al.* [12]), NdFe$_{1-x}$Mn$_x$AsO$_{0.89}$F$_{0.11}$ (Sato *et al.* [47]) with our PrFe$_{1-x}$Mn$_x$AsO$_{0.7}$F$_{0.3}$ samples (present work), while the AG model of pair-breaking for PrFe$_{1-x}$Mn$_x$AsO$_{0.7}$F$_{0.3}$ w.r.t La-1111 system is presented in the supplement Figure S3. The inset figure depicts the magnification of $T_c$ plot for La-1111 as a function of nominal Mn content ($x\%$) (**b**) Demonstration of the residual resistivity at 60 K as an increasing function of Mn concentrations ($x$).

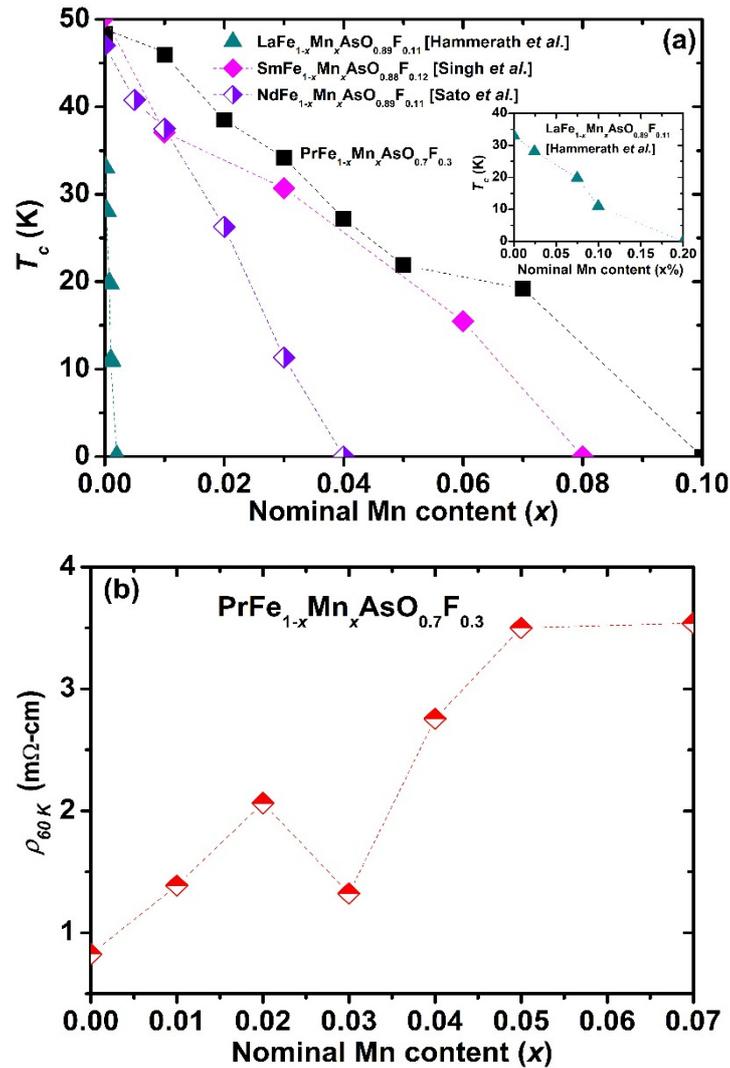